\documentclass{aastex}
\singlespace
\slugcomment{Accepted by the {\it Astrophysical Journal Letters}}
\def\la{\mathrel{\hbox{\rlap{\hbox{\lower4pt\hbox{$\sim$}}}\hbox{$<$}}}}
\def\ga{\mathrel{\hbox{\rlap{\hbox{\lower4pt\hbox{$\sim$}}}\hbox{$>$}}}}

\shortauthors{Park}
\shorttitle{SNR 1987A}

\begin{document}

\title{A New Evolutionary Phase of Supernova Remnant 1987A}

\author{Sangwook Park\altaffilmark{1}, Svetozar A. Zhekov\altaffilmark{2}, 
David N. Burrows\altaffilmark{3}, Judith L. Racusin\altaffilmark{4}, Daniel 
Dewey\altaffilmark{5} and Richard McCray\altaffilmark{6}} 

\altaffiltext{1}{Department of Physics, University of Texas at Arlington,
108 Science Hall, Box 19059, Arlington, TX 76019, USA; s.park@uta.edu} 
\altaffiltext{2}{Space and Solar-Terrestrial Research Institute, Moskovska str. 6,
Sofia 1000, Bulgaria}
\altaffiltext{3}{Department of Astronomy and Astrophysics, The Pennsylvania State
University, 525 Davey Laboratory, University Park, PA 16802, USA}
\altaffiltext{4}{NASA/Goddard Space Flight Center, 8800 Greenbelt Rd., Code 661, 
Greenbelt, MD 20771, USA}
\altaffiltext{5}{MIT Kavli Institute, Massachusetts Institute of Technology, 
Cambridge, MA 02139, USA}  
\altaffiltext{6}{JILA, University of Colorado, Box 440, Boulder, CO 80309, USA}


\begin{abstract}

We have been monitoring the supernova remnant (SNR) 1987A with {\it Chandra}
observations since 1999. Here we report on the latest change in the soft X-ray 
light curve of SNR 1987A. For the last $\sim$1.5 yr (since day $\sim$8000), 
the soft X-ray flux has significantly flattened, staying (within uncertainties)
at $f_{\rm X}$ $\sim$ 5.7 $\times$ 10$^{-12}$ erg cm$^{-2}$ s$^{-1}$ (corresponding 
to $L_{\rm X}$ $\sim$ 3.6 $\times$ 10$^{36}$ erg s$^{-1}$) in the 0.5--2 keV band. 
This remarkable change in the recent soft X-ray light curve suggests that the 
forward shock is now interacting with a decreasing density structure, after 
interacting with an increasing density gradient over $\sim$10 yr prior to day 
$\sim$8000. Possibilities may include the case that the shock is now propagating 
beyond a density peak of the inner ring. We briefly discuss some possible 
implications on the nature of the progenitor and the future prospects of our 
{\it Chandra} monitoring observations.         

\end{abstract}


\section {\label {sec:intro} INTRODUCTION}

The known distance ($d$ $\sim$ 50 kpc to the Large Magellanic Cloud 
[LMC]) and the identification of the progenitor (Sk --69$^{\circ}$202, 
a B3 I star) make SN 1987A a uniquely useful object for a detailed study of 
core-collapse supernova and the earliest stages of evolution of a supernova 
remnant (SNR). For the last $\sim$10 yr, SNR 1987A has been 
dominated by emission from the blast wave interacting with the dense 
circumstellar medium (CSM), dubbed ``the inner ring'', which was produced by 
interaction between the slow equatorial wind from the progenitor's red 
supergiant (RSG) phase and the fast spherical wind from the subsequent blue 
supergiant (BSG) phase \citep{bur95,lund91,luo91}. We have been monitoring 
SNR 1987A with {\it Chandra} since 1999. Thanks to the high angular resolution, 
our {\it Chandra} monitoring observations have allowed us to reveal 
unprecedented details of the dynamical and thermal evolution of the hot gas 
in SNR 1987A (e.g., Burrows et al. 2000; Michael et al. 2002; Park et al. 
2002,2004,2005,2006; Racusin et al. 2009). SNR 1987A has also been observed 
by {\it XMM-Newton} (e.g., Haberl et al. 2006; Sturm et al. 2010). 

The evolution of SNR 1987A has been highlighted by two characteristic events: 
(1) the emergence of optically-bright hot spots since 1997 (day $\sim$ 3700) 
and (2) the up-turn in the soft X-ray light curve that was coincident with 
a deceleration in the radial expansion rate in 2004 (day $\sim$ 6000). The 
former indicated the start of the interaction between the blast wave and 
the dense protrusions that had grown inward along the surface of the inner 
ring \citep{michael00}. The latter was interpreted as the event of the blast 
wave reaching the main body of the inner ring \citep{park05}. The continuous 
propagation of the blast wave into and/or beyond the inner ring would provide 
an excellent observational opportunity to study chemical and density structures 
of the progenitor's stellar winds and thus to reveal its detailed late-stage 
evolution.

Here we report the latest soft X-ray light curve from our {\it Chandra} 
monitoring observations. It is evident that the soft X-ray flux from SNR 1987A
has been nearly constant over the last $\sim$1.5 yr. This recent X-ray light 
curve is in contrast to the rapid brightening for the prior $\sim$10 yr. This 
remarkable change in the light curve indicates that SNR 1987A may now be 
entering yet another evolutionary phase. In this {\it Letter}, we report 
this new behavior in the soft X-ray light curve and briefly discuss 
possible implications. This work is based on ACIS and HETGS zeroth-order 
observations that we have performed as of 2010 September. The detailed 
analysis of the first-order HETGS spectrum obtained by our monitoring 
observations will be presented elsewhere.

\section{\label{sec:obs} OBSERVATIONS \& DATA REDUCTION}

As of 2010 September, we have performed a total of twenty-two monitoring 
observations of SNR 1987A with the Advanced CCD Imaging Spectrometer (ACIS) 
and the High Energy Transmission Gratings Spectrometer (HETGS) on board {\it 
Chandra}. To mitigate increasing photon pileup effects in the ACIS data due to 
the continuous brightening of SNR 1987A, we have switched our monitoring 
instrument configuration from the ACIS to the HETGS since 2008 July (day $\sim$ 
7800). The soft X-ray flux increase has been $\sim$10--15\% per $\sim$6-month, 
and the cross-calibration of the observed flux between these configurations 
is critical to make a smooth, reliable transition. Thus, we have performed 
{\it contemporaneous} observations using both the ACIS and HETGS (separated 
by $\la$ one week) for two epochs (2008 July and 2009 January). During the 
same time period, the correction tool for the charge transfer inefficiency 
(CTI) in the ACIS data has become available as part of the standard {\it 
Chandra} data reduction software CIAO. Thus, we adopted this standard 
CTI-correction instead of our previous method developed by Townsley et al. (2002). 
We carefully investigated the effect caused by these new instrument setup and 
CTI-correction method using the zeroth-order HETGS and bare ACIS data. The net 
effect is small ($\sim$3\% discrepancy in the observed 0.5--2 keV band flux, 
primarily caused by the new CTI-correction). This correction factor is accounted 
for in our updated X-ray light curves. We re-generated auxiliary response files 
for all twenty-two epochs using CALDB version 4.3.0. Other than these changes, 
we used the same data reduction process as that used in our previous work. 

In the spectral analysis to measure the X-ray fluxes, we used the non-equilibrium 
ionization (NEI) plane shock model \citep{bor01}. We implemented the NEI version 2 
in XSPEC with an augmented plasma model using the updated atomic data that account 
for the inner-shell lines\footnote{The updated model has been provided by K. 
Borkowski: cf. the discussion by Badenes et al. 2006.}. To be consistent with our 
previous works, we used a two-component NEI shock model to fit the observed X-ray 
spectra.  We varied the electron temperature, ionization timescale, and normalization 
for each of the soft and hard component in the model fits. We fixed elemental 
abundances for N, O, Ne, Mg, Si, S, and Fe at the best-fit values estimated by 
our deep {\it Chandra} grating data \citep{zhe06,zhe09}. We fixed the absorbing 
column at $N_{\rm H}$ = 2.35 $\times$ 10$^{21}$ cm$^{-2}$ based on our previous 
work \citep{park06}. Our {\it Chandra} monitoring observations are summarized 
in Table~\ref{tbl:tab1}. The observed X-ray fluxes for all {\it Chandra} 
monitoring observations are presented in Table~\ref{tbl:tab2}. We estimated the 
ACIS pileup effect based on the event grade distribution \citep{park07}. In 
Table~\ref{tbl:tab2}, the observed 0.5--2 keV band fluxes have been corrected 
for the pileup using the correction factor shown in Table~\ref{tbl:tab1}.

\section{\label{sec:lcs} Light Curve and Expansion Rate}

The latest X-ray light curves of SNR 1987A are shown in Figure~\ref{fig:fig1}a. 
As of 2010 September, the observed 0.5--2 keV band X-ray flux is $f_{\rm X}$ 
$\sim$ 57.9 $\times$ 10$^{-13}$ erg cm$^{-2}$ s$^{-1}$ ($L_{\rm X}$ $\sim$ 3.63 
$\times$ 10$^{36}$ erg s$^{-1}$ at $d$ = 50 kpc, after correcting for $N_{\rm H}$ 
= 2.35 $\times$ 10$^{21}$ cm$^{-2}$), which is $\sim$40 times brighter than it 
was in 1999 and nearly three orders of magnitude brighter than in 1992. While 
the X-ray flux has been significantly increasing over the last $\sim$20 yr, 
the latest data show that the X-ray flux increase rate has been significantly 
reduced. This is clearly seen in the soft band (0.5--2 keV) light curve 
(Figure~\ref{fig:fig1}a). For the last $\sim$1.5 yr, the 0.5--2 keV band X-ray 
flux has been nearly constant at $f_{\rm X}$ $\sim$ 57 $\times$ 10$^{-13}$ erg 
cm$^{-2}$ s$^{-1}$ ($L_{\rm X}$ $\sim$ 3.6 $\times$ 10$^{36}$ erg s$^{-1}$). 
This change is unclear in the 3--10 keV band, probably because of a lower 
flux increase rate and/or a physically different origin than the soft band 
emission. Upcoming monitoring observations are needed to reveal the nature 
of the hard band light curve. Nonetheless, the {\it flattening} of the soft 
X-ray light curve is compelling, and we discuss only the soft band light 
curve in this work.   

Figure~\ref{fig:fig1}b shows the updated radial expansion rate. The radius
of individual images (in the 0.3--8 keV band) and the expansion rate are 
estimated by the same methods described in Racusin et al. (2009). We tested 
any systematic discrepancy in the radius measurements between the ACIS and 
HETGS zeroth-order images using our cross-calibration data taken in 2008 July 
and 2009 January. We have also compared our radius estimates between the 
previous and current CTI corrections.  These calibration results showed 
negligible discrepancies in the radius estimates ($\la$1\%). Thus, for the 
data taken since 2008 July, we present the radial expansion rate based on 
the HETGS zeroth-order images with the standard CTI-correction method. The 
radial expansion rate has been consistent at $v$ $\sim$ 1600 km s$^{-1}$ 
for the last $\sim$6 yr.  

\section{\label{sec:disc} Discussion}


Ever since the blast wave started interacting with dense protrusions of the 
inner ring at day $\sim$3700, the soft X-ray light curve showed a rapid flux 
increase (e.g., Park et al. 2004). During days $\sim$6000--6500, the soft 
X-ray light curve turned up to show a steeper flux increase rate than before 
\citep{park05}. This phase was interpreted such that the evolution of SNR 
1987A entered a stage in which the shock is reaching the main body of the 
inner ring \citep{park05}. Such an interpretation was supported by the 
significant deceleration in the radial expansion rate at day $\sim$6000 
\citep{rac09}. Other observational supports included the substantial 
decrease in the line broadening measured by deep {\it Chandra} grating 
data \citep{zhe05,zhe09,dewey08}, the prevalence of optical spots 
all around the inner ring \citep{mccray05}, and the significant IR flux
increase \citep{bou06} at the same epoch. These evolutions of the X-ray 
light curve, radial expansion rate, and individual line widths suggested 
that X-ray emission was dominated by the blast wave interacting with the 
low-density H{\small II} region before the inner ring until day $\sim$4000, 
and that since then soft X-rays primarily originate from the blast wave 
interacting with the dense inner ring. Recently, line fluxes from some 
individual optical spots showed a hint for flattening of the optical light 
curve at day $\sim$7000 \citep{gron08}.

\begin{figure}[]
\figurenum{1}
\centerline{\includegraphics[angle=0,width=0.85\textwidth]{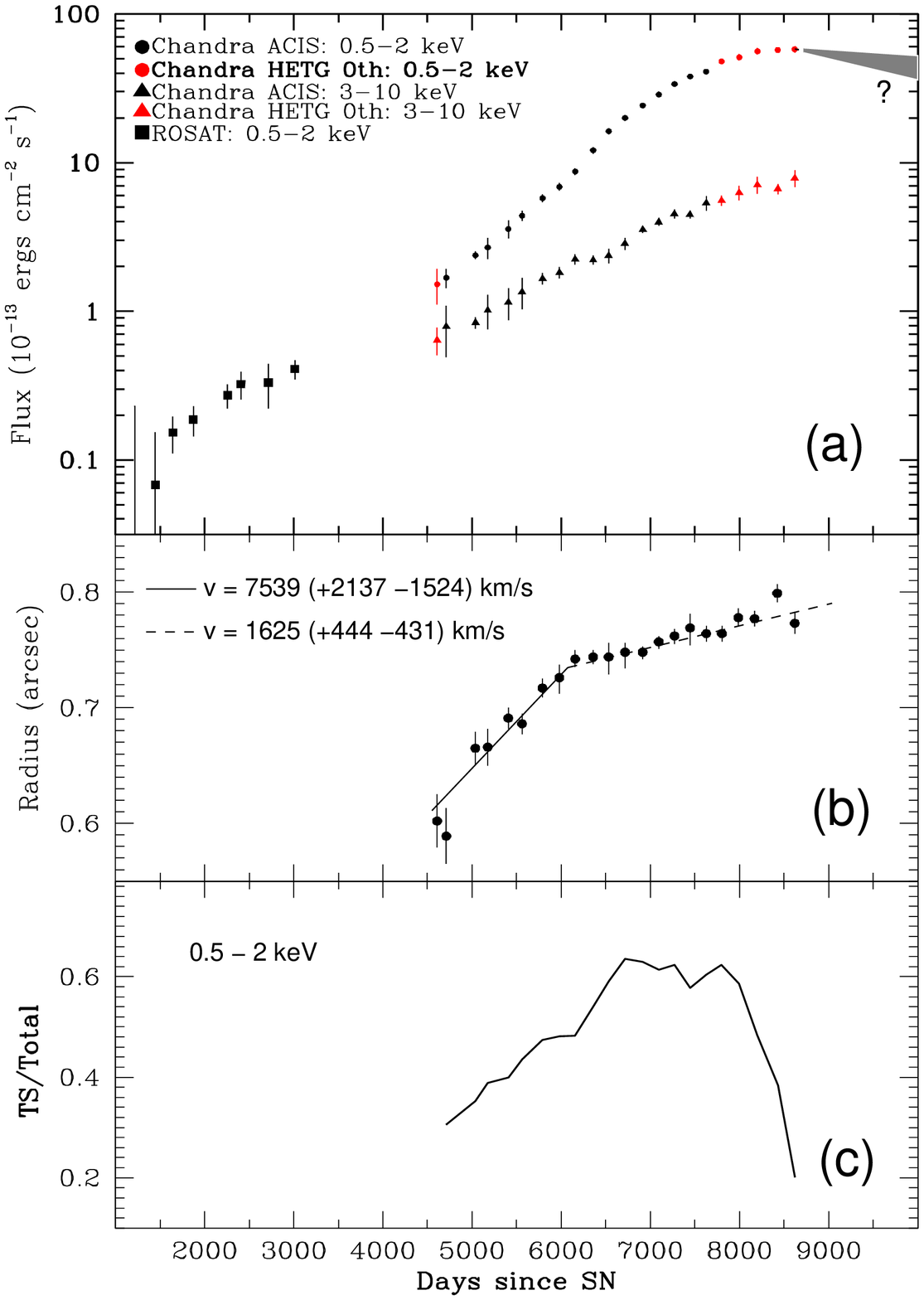}}
\figcaption[]{(a) X-ray light curves of SNR 1987A from our {\it Chandra} monitoring
observations. The {\it ROSAT} data are taken from Hasinger et al. (1996). The shaded 
area shows the range of the predicted light curve assuming that $f_{\rm X}$ $\propto$ 
$t^{-1}$--$t^{-3}$ starting at day $\sim$8600 (see the text). (b) Radial expansion 
measure from the fits to the quadrilateral lobes plus torus model for the broken 
linear fit with an early fast shock velocity (solid line) transitioning to the later 
slower shock velocity (dashed line). (c) The temporal evolution of the fractional 
flux from the transmitted shock component as estimated by the RSS model fits by Zhekov 
et al. (2010).
\label{fig:fig1}}
\end{figure}

Our {\it Chandra} monitoring observations show that the flux increase rate 
appears to be lower at days $\sim$6500--8000 than at days $\sim$6000--6500 
(i.e., $f_{\rm X}$ $\sim$ $t^{6}$ as opposed to $\sim$ $t^{10}$ for 
days $\sim$6000--6500). Subsequently, the soft X-ray flux has been nearly 
constant for $\sim$1.5 yr (Figure~\ref{fig:fig1}a). The latest flattening 
of the X-ray light curve indicates that SNR 1987A is now entering another 
phase. Assuming that the observed X-ray flux is dominated by the 
interacting density, the X-ray flux can be considered to be proportional 
to the volume emission measure $EM$ $\sim$ $n_e^2$ $V$, where $n_e$ is the 
electron density and $V$ is the emission volume (Although there was a 
deceleration in the expansion rate at day $\sim$6000, the shock velocity 
has been constant since then. Also, the temperature dependence of the X-ray 
flux is small: i.e., $f_{\rm X}$ $\propto$ $T^{-0.6}$ \citep{mckee77}). 
The exact geometry of the X-ray emitting volume in SNR 1987A is uncertain. 
Previous works suggested that the bulk of X-ray emission 
originates from a disk-like shell between the forward and reverse shocks, 
which encompasses the inner ring. An {\it upper limit} for such an emitting 
volume may be approximated by a spherical shell with an expanding thickness, 
thus $V$ $\propto$ $r^3$ where $r$ is the SNR radius. A {\it lower limit} 
of the X-ray emitting volume could be a circular disk with a small thickness, 
thus $V$ $\propto$ $r$.  These two cases can serve as conservative limits 
that are useful to approximate the temporal evolution of the X-ray 
emitting volume. The reality is likely in between these limits. For $V$ 
$\propto$ $r^s$ where $s$ = 1--3, the X-ray flux would be $f_{\rm X}$ 
$\propto$ $EM$ $\propto$ $n_e^{2}$ $r^s$. Assuming that $n_e$ $\propto$ 
$r^{\alpha}$, we have $f_{\rm X}$ $\propto$ $r^{2{\alpha}+s}$. If we 
fit the light curve as a power law (PL), $f_{\rm X}$ $\propto$ 
$t^{\beta}$, then since the expansion is linear (Figure~\ref{fig:fig1}b), 
$f_{\rm X}$ $\propto$ $r^{\beta}$, and therefore $\beta$ = 2$\alpha$ + $s$. 
This simple relation suggests that, for $\beta$ $\sim$ 6 during day $\sim$ 
6500--8000, the blast wave was still interacting with a radially-increasing 
density ($n_e$ $\propto$ $r^{1.5}$ for $s$ = 3, and $\propto$ $r^{2.5}$ for 
$s$ = 1), but at a lower rate than that ($n_e$ $\propto$ $r^{3.5}$ for 
$s$ = 3, and $\propto$ $r^{4.5}$ for $s$ = 1) during day $\sim$ 6000--6500. 
Since day $\sim$8000 the flux has been constant, which corresponds to 
the shock interacting with a decreasing radial density profile ($n_e$ 
$\propto$ $r^{-1.5}$ for $s$ = 3, and $\propto$ $r^{-0.5}$ for $s$ = 1). 
We note that our PL fit analysis is likely an oversimplification. We intend 
to provide insights on physical implications imposed by the recent flattening
of the light curve. Quantitative interpretations inferred from these simple 
PL fits before day $\sim$8000 are far from conclusive. 

These simple estimates of the density structure into which the blast wave 
is now propagating suggest that a significant fraction of X-ray emission 
is now coming from the shock interacting with a CSM having lower density 
than the densest region along the inner ring. Possible origins for such 
a change in the ambient density may include an intriguing scenario that the 
blast wave might have reached a density peak of the inner ring at day 
$\sim$8000, and that it is now propagating into a decreasing radial density 
beyond the inner ring. A likely scenario for the progenitor's evolution before 
the explosion of SN 1987A is a BSG phase that had gone through an earlier RSG 
stage. The standard stellar evolutionary models predict an RSG wind with a 
constant velocity and a radially-decreasing density profile ($n$ $\sim$ 
$r^{-2}$). Our estimates of the latest radial density profile ($n$ $\sim$ 
$r^{-1.5}$ - $r^{-0.5}$) suggest that the blast wave of SNR 1987A may be 
entering the RSG wind beyond the interacting layer between the BSG and RSG 
winds. If this is true, the soft X-ray light curve would eventually decrease 
at a rate of $f_{\rm X}$ $\propto$ $t^{-1}$ - $t^{-3}$ (depending on $s$) 
as the shock would be interacting with the RSG wind ($n$ $\sim$ $r^{-2}$). 
If the overall RSG wind is spherical beyond the inner ring, $f_{\rm X}$ 
$\propto$ $t^{-1}$ is expected, which is generally consistent with standard 
SN models \citep{chev82}.

We note that Borkowski et al. (1997) predicted the soft X-ray light curve 
of SNR 1987A based on 2-dimensional hydrodynamic simulations. In their models, 
the soft X-rays are dominated by emission from dense gas that has been heated 
by shocks transmitted into the CSM ring as it is overtaken by the blast wave. 
The resulting soft X-ray light curves are sensitive to the assumed gas density, 
$n_0$, in the inner ring. The X-ray flux in their model for $n_0$ = 32000 amu 
cm$^{-3}$ reaches a peak value of $f_{\rm X}$ = 6 $\times$ 10$^{-12}$ erg 
cm$^{-2}$ s$^{-1}$ at $t$ = 17 yr, which roughly agrees with the observed 
results shown in Figure~\ref{fig:fig1}a ($f_{\rm X}$ = 5.7 $\times$ 10$^{-12}$ 
erg cm$^{-2}$ s$^{-1}$ at $t$ = 23 yr). However, the actual hydrodynamics of 
the interaction are more complex than such simple simulations. While Borkowski 
et al. (1997) assumed that the CSM ring was a circular torus of constant 
density, the inner ring is not a circular torus, having many inward protrusions 
manifested by the optical spots (e.g., Sugerman et al. 2002). The density in 
the inner ring spans a range of a factor of $\sim$30 \citep{mat10}. Nonetheless, 
these results suggest that more realistic 3-dimensional simulations for the 
interaction of the blast wave and the inner ring would help to reveal the 
detailed nature of the X-ray light curve.   

Other models for SN 1987A include cases that the progenitor might have
gone through a Luminous Blue Variable (LBV) phase \citep{smith07}, which 
most likely involved episodic eruptions of strong asymmetric stellar 
winds. In this scenario, the inner ring might have been the relic dense 
CSM produced by the latest LBV eruption before the SN. The radial density 
profile beyond the inner ring then might have a periodic ``density wave''. 
Such a density structure may result in a periodic variation in the soft 
X-ray light curve in the future. 

We also consider the latest soft X-ray light curve within the frame work 
of our reflected shock structure (RSS) model \citep{zhe10}. The current 
radial expansion rate appears to be well-fitted by the RSS model, as the 
expansion rate stays the same until day $\sim$8600. The RSS model fit shows 
that the contribution in the soft X-ray emission from the transmitted shock 
into the densest region of the inner ring peaks at day $\sim$6700--7800, and 
then it significantly decreases after day $\sim$8000 (Figure~\ref{fig:fig1}c). 
These results generally support that soft X-ray emission originating from 
a low density region has recently become significant.

Although the scatter in the radius measurements appears to become large 
for the last $\sim$2 yr, we do not see clear evidence for an acceleration 
in the expansion as of 2010 September (Figure~\ref{fig:fig1}b). This is 
probably because the shock is in transition between the inner ring and 
the RSG wind, and our data may not yet be sensitive to an accelerated 
expansion rate. Or, the shock may still be interacting with the inner 
ring (see above discussion). Future {\it Chandra} observations are 
essential to monitor any change in the expansion rate as well as 
in X-ray light curves. We note that the X-ray radius of SNR 1987A 
($\sim$0$\farcs$78 at day $\sim$8600) is slightly smaller than that 
of the optical inner ring ($\sim$0$\farcs$83, Jakobsen et al. [1991]). 
Our radius estimates are based on a simple image model (assuming 
4-lobes and an underlying torus) fit to deconvolved images to measure the 
radius of the peak intensity of the SNR \citep{rac09}. Other methods 
have been used to measure the size of the SNR in optical \citep{jakob91} 
and radio bands \citep{ng09}. Thus, it is not surprising to see systematic 
offsets among the radius measurements in different wavelengths. Ng et al. 
(2009) confirmed that there is a systematic discrepancy (by $\sim$20 \%) 
between the radio and X-ray measurements of the radius of SNR 1987A because 
of the different methods used for X-ray and radio data. Therefore, direct 
comparisons of the SNR's angular size between different wavelength bands 
are difficult.

Our continuing {\it Chandra} monitoring observations of the X-ray flux and 
the expansion rate are essential to reveal the true origin for the dynamically 
changing soft X-ray light curve. The high resolution first-order HETGS data 
of our monitoring observations will be useful to reveal the spectral evolution 
of SNR 1987A. Particularly, our deep HETGS observation of SNR 1987A that is 
scheduled in 2011 March will be critical to study the detailed evolution of 
individual line broadenings and flux ratios, and thus the SNR's evolutionary 
phase. If the contribution from the reverse shock in the observed X-ray 
emission has substantially increased, we may expect a bleaching-out of optical 
Ly$_{\alpha}$ and H$_{\alpha}$ lines in near future due to photo-ionization 
(by X-ray photons) of neutral H just ahead of the reverse shock as predicted 
by Smith et al. (2005).

\acknowledgments

The authors thank K. Borkowski for providing the augmented NEI shock model. 
This work was supported in part by SAO under {\it Chandra} grants GO9-0082X 
and GO0-11072X.

\begin{deluxetable}{ccccccc}
\footnotesize
\tablecaption{{\it Chandra} Observations of SNR 1987A.
\label{tbl:tab1}}
\tablewidth{0pt}
\tablehead{\colhead{ObsID} & \colhead{Date} & \colhead{Age\tablenotemark{a}} & 
\colhead{Instrument\tablenotemark{b}} & \colhead{Exposure} & \colhead{Counts} & 
\colhead{Correction\tablenotemark{c}} \\ 
&  &  &  & \colhead{(ks)} & & }   
\startdata
124, 1387 & 1999 October & 4609 & HETG (3.1 s) & 116.1 & 690 & 1.02 \\
122 & 2000 January & 4711 & ACIS-S3 (3.2 s) & 8.6 & 607 & 1.05 \\
1967 & 2000 December & 5038 &  ACIS-S3 (3.2 s) & 98.8 & 9030 & 1.04 \\
1044 & 2001 April & 5176 & ACIS-S3 (3.2 s) & 17.8 & 1800 & 1.04 \\
2831 & 2001 December & 5407 & ACIS-S3 (3.1 s) & 49.4 & 6226 & 1.06 \\
2832 & 2002 May & 5561 &   ACIS-S3 (3.1 s) & 44.3 & 6427 & 1.10 \\
3829 & 2002 December & 5791 & ACIS-S3 (3.1 s) & 49.0 & 9277 & 1.08 \\
3830 & 2003 July & 5980 & ACIS-S3 (3.1 s) & 45.3 & 9668 & 1.12 \\
4614 & 2004 January & 6157 &  ACIS-S3 (3.1 s) & 46.5 & 11856 & 1.18 \\
4615 & 2004 July & 6359 & ACIS-S3 (1.5 s) & 48.8 & 17979 & 1.10 \\
5579, 6178 & 2005 January & 6533 & ACIS-S3 (0.4 s) & 48.3 & 24939 & 1.02 \\
5580, 6345 & 2005 July & 6716 & ACIS-S3 (0.4 s) & 44.1 & 27048 & 1.04 \\
6668 & 2006 January & 6914 & ACIS-S3 (0.4 s) & 42.3 & 30940 & 1.04 \\
6669 & 2006 July & 7095 & ACIS-S3 (0.4 s) & 36.4 & 30870 & 1.05 \\
7636 & 2007 January & 7271 & ACIS-S3 (0.4 s) & 33.5 & 32798 & 1.06 \\
7637 & 2007 July & 7446 & ACIS-S3 (0.4 s) & 25.7 & 27945 & 1.07 \\
9142, 9806 & 2008 January & 7626 & ACIS-S3 (0.2 s) & 9.3 & 12008 & 1.04 \\
9144 & 2008 July\tablenotemark{d} & 7799\tablenotemark{d} & 
HETG\tablenotemark{d} (1.1 s) & 42.0 & 5174 & 1.03 \\
10852, 10221 & 2009 January\tablenotemark{d} & 7997\tablenotemark{d} & 
HETG\tablenotemark{d} (1.1 s) & 71.5 & 9447 & 1.03 \\
10853, 10854 & & & & & & \\
10855 & & & & & & \\
10222, 10926 & 2009 July/ & 8202 & HETG (1.1 s) & 58.2 & 8505 & 1.03 \\
             & September     & & & & & \\
12125, 12126 & 2010 March & 8429 & HETG (1.0 s) & 63.9 & 9332 & 1.03 \\
11090 & & & & & & \\
13131, 11091 & 2010 September & 8619 & HETG (1.0 s) & 54.4 & 8361 & 1.03 \\
\enddata

\tablenotetext{a}{Days since the SN explosion.}
\tablenotetext{b}{For HETG, the zeroth-order data are used. The frame-time
is shown in parentheses.}
\tablenotetext{c}{This is the flux correction factor accounting for
the CCD pileup and cross-calibration with our previous CTI-correction
method (by Townsley et al.) that was adopted before 2008 July.}
\tablenotetext{d}{For this epoch, simultaneous observations with the ACIS
and HETG were performed to calibrate fluxes between two detectors. We here
present only HETG data.}

\end{deluxetable}

\clearpage

\begin{deluxetable}{ccc}
\footnotesize
\tablecaption{The Observed {\it Chandra} X-ray flux of SNR 1987A.
\label{tbl:tab2}}
\tablewidth{0pt}
\tablehead{\colhead{Age\tablenotemark{a}} & \colhead{Flux\tablenotemark{b}} 
& \colhead{Flux\tablenotemark{b}} \\ 
\colhead{(days)} & \colhead{(0.5--2 keV)} &  \colhead{(3--10 keV)} }   
\startdata
4609 & 1.52$\pm$0.41 & 0.64$\pm$0.13 \\
4711 & 1.68$\pm$0.25 & 0.79$\pm$0.30 \\
5038 & 2.38$\pm$0.14 & 0.84$\pm$0.08 \\ 
5176 & 2.68$\pm$0.43 & 1.02$\pm$0.27 \\ 
5407 & 3.57$\pm$0.50 & 1.15$\pm$0.28 \\
5561 & 4.39$\pm$0.35 & 1.35$\pm$0.33 \\
5791 & 5.77$\pm$0.35 & 1.66$\pm$0.15 \\
5980 & 6.88$\pm$0.41 & 1.82$\pm$0.16 \\
6157 & 8.70$\pm$0.44 & 2.24$\pm$0.18 \\
6359 & 12.11$\pm$0.61 & 2.21$\pm$0.16 \\
6533 & 16.26$\pm$0.81 & 2.37$\pm$0.26 \\
6716 & 19.95$\pm$0.80 & 2.85$\pm$0.26 \\
6914 & 24.23$\pm$0.73 & 3.54$\pm$0.18 \\
7095 & 28.72$\pm$0.86 & 3.98$\pm$0.24 \\
7271 & 33.76$\pm$0.68 & 4.52$\pm$0.32 \\
7446 & 37.91$\pm$1.14 & 4.46$\pm$0.27 \\
7626 & 40.89$\pm$1.64 & 5.36$\pm$0.59 \\
7799 & 47.93$\pm$1.92 & 5.55$\pm$0.44 \\
7997 & 51.10$\pm$2.56 & 6.28$\pm$0.69 \\
8202 & 56.07$\pm$2.80 & 7.10$\pm$0.92 \\
8429 & 57.25$\pm$2.30 & 6.66$\pm$0.53 \\
8619 & 57.89$\pm$2.90 & 7.88$\pm$1.02 \\
\enddata

\tablenotetext{a}{Days since the SN explosion.}
\tablenotetext{b}{Observed X-ray fluxes are in units of 10$^{-13}$ 
erg cm$^{-2}$ s$^{-1}$. Errors are with 90\% confidence level, 
estimated by ``flux error'' command in XSPEC.}

\end{deluxetable}

\clearpage

\end{document}